\documentclass[a4paper]{article}

\usepackage{INTERSPEECH2022}
\usepackage{svg}
\usepackage{caption}
\usepackage{subcaption}
\usepackage{multicol}
\usepackage{multirow}
\usepackage{amsthm}
\usepackage{enumitem}

\usepackage{hyperref}
        \usepackage{setspace}

\hypersetup{
  colorlinks,
  allcolors=.,
  urlcolor=blue,
}

\title{A Scalable Model Specialization Framework for Training and Inference using Submodels and its Application to Speech Model Personalization}
\name{Fadi Biadsy, Youzheng Chen, Xia Zhang, Oleg Rybakov, Andrew Rosenberg, Pedro J. Moreno}
\address{Google LLC}
\email{\{biadsy,josephychen,xiaz,rybakov,rosenberg,pedro\}@google.com}

\begin{document}

\setstretch{0.93}

\maketitle
\begin{abstract}
Model fine-tuning and adaptation have become a common approach for model specialization for downstream tasks or domains. Fine-tuning the entire model or a subset of the parameters using light-weight adaptation has shown considerable success across different specialization tasks. Fine-tuning a model for a large number of domains typically requires starting a new training job for every domain posing scaling limitations. Once these models are trained, deploying them also poses significant scalability challenges for inference for real-time applications. In this paper, building upon prior light-weight adaptation techniques, we propose a modular framework that enables us to substantially improve scalability for model training and inference. We introduce {\em Submodels} that can be quickly and dynamically loaded for on-the-fly inference. We also propose multiple approaches for training those Submodels in parallel using an embedding space in the same training job. We test our framework on an extreme use-case which is speech model personalization for atypical speech, requiring a Submodel for each user. We obtain $128\times$ Submodel throughput with a fixed computation budget without a loss of accuracy. We also show that learning a speaker-embedding space can scale further and reduce the amount of personalization training data required per speaker.
\end{abstract}


\section{Introduction}
Neural model fine-tuning and adaptation have become a standard approach for model specialization or task customization.
Fine-tuning the entire model or a subset of the parameters of the model has shown substantial gains for a wide range of downstream tasks (e.g.,~\cite{Brown2020,Bommasani2021Fundation,Shor2019,biadsy2019parrotron,Gururangan2020,Yue2021}).  However, these techniques pose scalability limitations for {\em both} training and inference when applied to a {\em large} number of tasks or use-cases within a task (such as, domains, conditions, or even speakers).


For any number of specializations, the cost scales linearly with each task or use-case. When this number is large from, say, personalizing a separate speech model for each user, or customizing a model to contextual domains, this is unfeasible.  There are similar scaling problems to serve or deploy a large number of customized models whether on-server or locally on-device for each task or use-case. 

Light-weight adaptation techniques, such as \cite{Shen2021LowRank, Zaken2019biases,Lester2021,houlsby2019parameter,bapna2019simple,tomanek2021residual,AdapterFusionPfeiffer2020,multitaskAdapters} have shown significant reduction on memory, computation and storage costs by learning or adapting significantly less parameters than that of the full model.
We are inspired by this work to propose a framework that scales both training and serving for many use-cases. We evaluate our approach on speech model personalization, requiring one model for each user. We demonstrate this framework on personalizing Parrotron for atypical speech conversion and recognition~\cite{biadsy2019parrotron}.

Prior work has shown that fine-tuning the entire or a large subset of the parameter space can successfully personalize a speech model for end-to-end Automatic Speech Recognition (ASR) and speech conversion~\cite{Zhu2019,Shor2019,Gale2019,mustafa2014,biadsy2019parrotron,chen2021conformer,green2021,kannan2019large}. 
To motivate our framework, we summarize the limitations of employing model fine-tuning for speech model personalization:
\begin{itemize}[leftmargin=*]
  \item \textbf{Inference:} Using a fine-tuned model requires loading all parameters per request on the server, for all users.
  Loading a large model from disk may take significant time per request which prevents scaling. Alternately, keeping many models in memory will not scale due to limited RAM capacity and cost. 
  \item \textbf{Training:} It is expensive to personalize an entire model for every user.
  Training such a model requires starting a new training job which may take hours to complete and will result into a separate large model  for every user. 
  \item \textbf{Data requirements:} Fine-tuning all model parameters on a small amount of data may result on over-fitting and catastrophic forgetting~\cite{french999128}.
  \item \textbf{On-Device local models}: Each of these limitations are more severe for on-device specialization, where resources are more limited and transfer bandwidth is slower and more expensive. 
  
\end{itemize}

This paper makes the following contributions: (1)  We lay out a modular framework that serves as a clarifying perspective on how to approach the model specialization problem, specifically addressing scaling issues for training and serving;  (2) We propose dynamic loading of Submodels on-demand to perform on-the-fly inference; (3) We show how this framework enables an effective personalization design for speech models, specifically Parrotron; and (4) We propose a novel approach to Submodeling by learning a shared embedding space which minimizes the personalization data requirements for new speakers.

\section{Framework: Basemodels and Submodels}

To formalize our modular design, we define two components: {\bf Basemodel} and {\bf Submodels}.
The focus of this paper is a model trained to perform a single task and related parallel tasks.
We note, however, that there is no limitation in our design for modeling and customizing to less related tasks with distinct inputs or outputs. 
We also focus on real-time scalability requirements.   

\subsection{Basemodel}

A Basemodel is a fully-trained model capable of doing inference for the desired task, trained on a general distribution of the data (also proposed by~\cite{multitaskAdapters} and others). A Basemodel is not specialized and would work well on the traffic similar to the training distribution without any further modification or customization. 
Note that this is in contrast to Foundation models which are typically intermediary assets, and {\em require} adaptation to perform inference.~\cite{Bommasani2021Fundation}
A speaker- and domain- independent ASR model is a good example of a Basemodel since it can be used as is to serve most users or domains with acceptable accuracy and it's not specialized for a given domain. 

\subsection{Submodels}
\label{sec:submodels}

A Submodel is a copy of a subset of parameters of the Basemodel, or extra new parameters added to the Basemodel that can be specialized or adapted for a given use-case while freezing the remaining Basemodel parameters.  Submodels are used to bias the
Basemodel towards a particular use-case.

{\bf Activation} Submodels must have an ability to be activated and deactivated dynamically during inference without the need to reload the Basemodel. Deactivating a Submodel during inference will result in a model equivalent to the Basemodel.
Any component that cannot be easily activated or deactivated during inference cannot act as a Submodel due to accelerator compilation and optimization requirements, which may take significant amount of time for real-time applications. 

{\bf Size} A Submodel must be sufficiently small that can be {\em quickly} loaded on-demand and dynamically from storage and easily enabled in the Basemodel during inference. 



{\bf Data Independence} Submodels ensure that data relevant for a specific use-case can be used independently. Improving on one use-case guarantees, by design, to not impact Submodels of other use-cases -- i.e., independent data distributions.  Submodels can be added, updated, and even shipped at any time. 

Finally, such a modular design may enable the research community to focus on exploring the internal structures of Basemodel and Submodels separately, as long as a clear `protocol' is defined. In this paper, we also show that Submodels can also be parameterized by a side input to make them more flexible and scalable. Some examples of Submodels include the following: a subset of layers of the encoder or decoder, residual adapters added to the encoder or decoder for ASR.~\cite{houlsby2019parameter}; layer biases of the entire model or encoder~\cite{Zaken2019biases}; adapted layer amplitudes \cite{Swietojanski2024}, or the joint network of a transducer ASR model.

In this paper, we use speech model personalization as an example of the proposed framework. We choose residual adapter layers~\cite{houlsby2019parameter} as our choice of Submodels for several reasons: (1) Adapter layers can easily be added to the encoder which is generally responsible for modeling the acoustic-phonetic of speech from the acoustic signal, 
 (2) Due to their residual connection one can disable the Submodel by simply setting the residual factor to zero, reverting the model to the Basemodel; (3) The size of this Submodel can be easily controlled by a bottleneck dimension;
(4) Controlling the bottleneck dimension is internal to the Submodel, allowing the use of a pre-compiled and optimized execution graph for fast inference while being able to replace the tensors shape dynamically; 
 (5) We have seen previously that adapters successfully model atypical and accented speech for ASR personalization and specialization.~\cite{tomanek2021residual}


\section{Model and Task}

In this section, we present the Basemodel and data we use for our experiments and show results comparing the unadapted Basemodel and the fully fine-tuned model for model personalization. The fine-tuned model per user can be viewed as our `top line' we aim to reach. However, this per-user fine-tuning approach is prohibitively expensive (Section~\ref{sec:scaling}).




\begin{figure}[ht]
\vspace{-5mm}
  \centering
  \includegraphics[width=\linewidth]{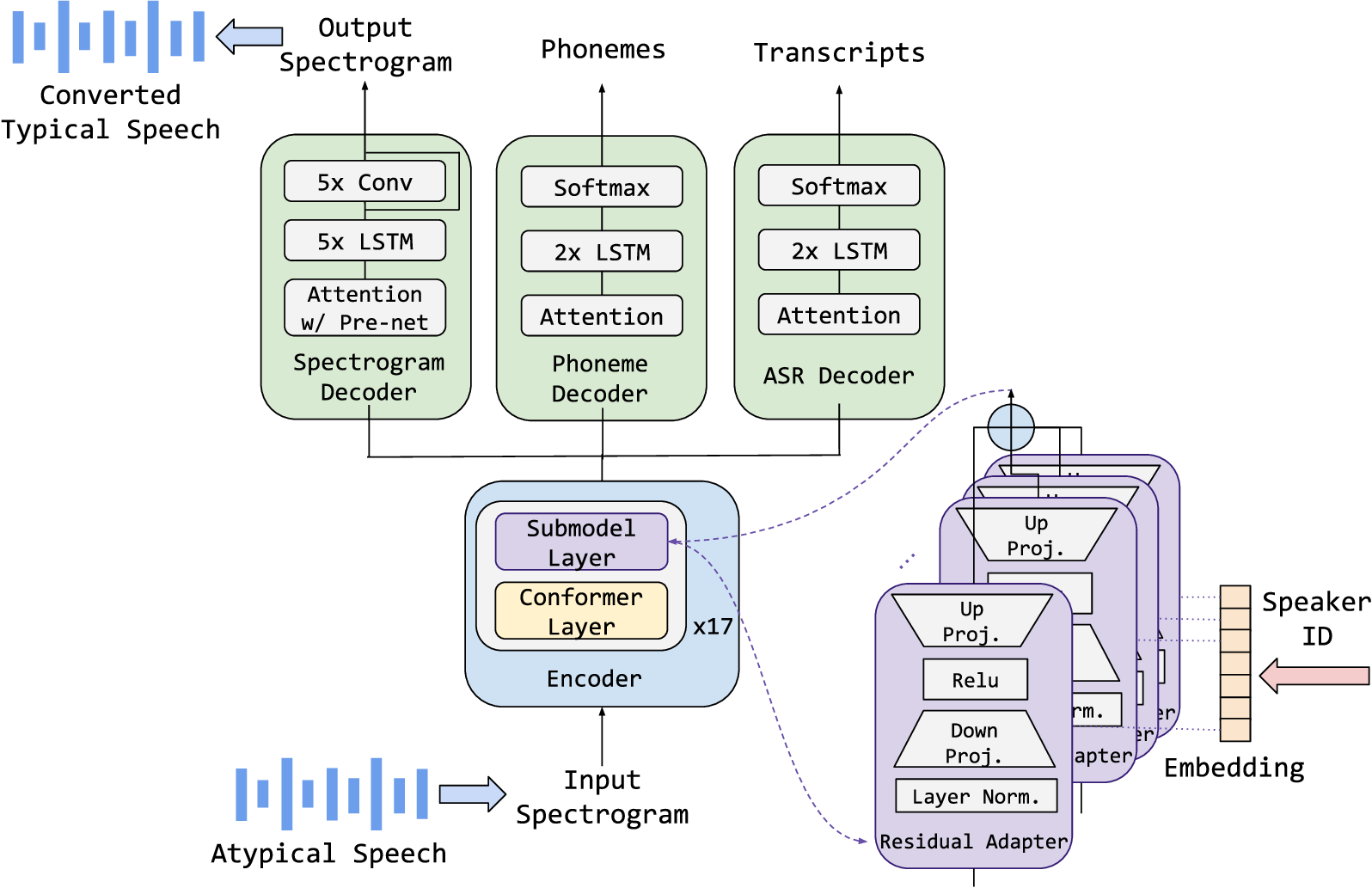}
  \caption{Parrotron Model with Embedding Submodel}
    \label{fig:parrotronWithEmbedding}
\vspace{-6mm}
\end{figure}
\subsection{Parrotron Basemodel}

Parrotron is a speech-to-speech conversion model that can be personalized to convert dysarthric and atypical speech into a synthesized typical and fluent speech (\href{https://bit.ly/37wDb1g}{demo}).
The Parrotron model is trained end-to-end to directly map the input spectrogram to another spectrogram, without using an intermediate discrete representation. The generated output spectrogram is then passed to a vocoder to produce a time-domain waveform.  The Parrotron model has been extended to also produce transcripts in parallel, effectively making it an ASR system as well.~\cite{doshi2021extending} 

Parrotron is initially trained on typical
speech  to obtain a many-to-one speech
conversion, yielding a Basemodel.
It has been shown that fine-tuning the full model on
hundreds of speech utterances from a single speaker obtains a one-to-one speech
conversion model to convert from atypical to typical speech with significant reduction in Word Error Rate (WER) and MOS scores when compared to the  Basemodel~\cite{biadsy2019parrotron,doshi2021extending,chen2021conformer}. 


\vspace{-2mm}
\subsection{Data}
Our data is a subset of the Euphonia corpus~\cite{macdonald2021disordered}. It consists of 128 speakers with a variety of speech impairments: 37 speakers with ALS, 24 Down-Syndrome, 21 Cerebral Palsy, 8 Parkinson's, 5 hearing implements, 5 Unknown, 4 Multiple Sclerosis, 2 Stroke, and the remaining 22 of other etiologies. Speech pathologists were asked to rate the severity of impairments: 56 speakers were rated as Mild, 40 Moderate, 29 Severe, and 3 Typical speech. These speakers were presented with prompts and asked to read them out-loud. Prompts are from a variety of domains, such as home automation, caregiver phrases, and conversational sentences. The mean, median, and standard deviation of the number of utterances per speaker is 2154, 1769 and 1399, respectively (min:  250, max: 4250). For every speaker, we hold 80\% for train, 10\% for dev and 10\% for test.  

\subsection{Basemodel vs. Model Fine-tuning Results}
\label{sec:Submodel}

As shown in Table~\ref{tab:finetuning_vs_Submodel}, running Google's ASR engine, our 128 speakers results in a 50.5\% average WER and a median of 50\%. This large average WER is not surprising. While the ASR engine was trained on thousands of hours of anonymized speech data, all training data are from speakers with {\em typical} speech.  Differences in etiologies and severity explain the large standard deviation across speakers.

Using a many-to-one, unadapted, Parrotron Basemodel trained on the same data as the Google's ASR engine, we also report the average WER across 128 dysarthric speakers.  Parrotron produces {\em two} outputs, in parallel, given an input utterance: converted speech output from the Spectrogram decoder, and an automatic transcript output from the ASR decoder. 
To evaluate the correctness of the converted speech, we pass the vocoded Parrotron spectrogram output to Google's ASR engine and report WERs.  The conversion WERs correlate strongly with the WERs of the automatic transcripts obtained from Parrotron's ASR decoder ($r=0.9993, RMSE=3.3$), 
%
across all experiments. For simplicity, we only report conversion WERs.


The Parrotron unadapted Basemodel achieves an average WER of 48.4\% across speakers,  somewhat better than that obtained by a state-of-the-art ASR engine (Table~\ref{tab:finetuning_vs_Submodel}). 
%
We find that fine-tuning the entire model (i.e., adapting all parameters of the Basemodel including encoder and decoders) for each speaker exerts substantial improvements across all our speakers with an average WER of 14.2\%. This is consistent with our previous work that Parrotron model fine-tuning achieves high quality model personalization for atypical speech~\cite{biadsy2019parrotron, doshi2021extending,chen2021conformer, tomanek2021residual}. For this experiment, we employ the same model architecture and the same fine-tuning procedure described in~\cite{chen2021conformer}. The current work confirms that a fine-tuning strategy performs well on speakers across etiologies and severities. For this and all other experiments we use a batch size of 8 and train for 30K steps.

\begin{table}[t]
\scriptsize
    \centering
    \caption{Comparing approaches (WER) - 128 atypical speakers.}
    \begin{tabular}{cl|rrr}
        
        & Approach & Mean WER & Median & SD \\
        \hline
        &{Google's ASR Engine} & 50.51 & 49.95 & 30.53 \\
        \hline
        & Basemodel           & 48.43 & 48.29 & 29.08 \\
        & Full model fine-tuning    & 14.23 &  9.98 & 12.00 \\
        & Speaker Submodel    & 15.27 & 10.63 & 12.29 \\
        & Pooling Submodel    & 22.64 & 18.60 & 14.82 \\
        & One-hot-embedding Submodel   & 15.17 & 10.45 & 8.57 \\
        & Real-embedding Submodel    & 18.54 & 12.93 & 14.16 \\
        \hline
        
    \end{tabular}
    \label{tab:finetuning_vs_Submodel}
    \vspace{-5mm}
\end{table}

\vspace{-0.14cm}
\section{Scaling Inference and Training}

\label{sec:scaling}

\subsection{Parrotron Submodel} 
Our choice of Submodel in this paper is based on residual adapters. Our Submodel consists of 17 residual adapter blocks, each added between each layer of the 17-layer conformer encoder in the Parrotron Basemodel~\cite{houlsby2019parameter,bapna2019simple,tomanek2021residual,conformer}. Each adapter block starts with layer normalization, followed by a down-projection to a bottleneck dimension $d_b$ followed by $Relu$, and finally another feed-forward layer with up-projection to the original input dimension, as shown in Figure~\ref{fig:parrotronWithEmbedding}.
The bottleneck dimension $d_b$ enables us to control the number of parameters of each adapter. In all experiments, unless stated otherwise, we use a bottleneck of $d_b=64$. All weights are randomly initialized.

Instead of fine-tuning the entire model, we adapt only the Submodel (i.e., all residual adapters) while freezing all parameters of the Basemodel. As shown in Table~\ref{tab:finetuning_vs_Submodel}, we observe that Submodel adaptation results in a WER that is about 1\% (absolute) worse than that of a fully fine-tuning model (statistically significant according to a paired t-test with $p < .001$). 
%
`However, given the substantial scaling benefits, and improvement over the Basemodel, we find that the use of residual adapters is an effective approach to personalizing speech conversion models. The fully fine-tuned model per speaker requires hosting 165M parameters (of 668 MB on disk) per user, whereas the adapted Submodel has only 1.2M parameters (of 4.6MB on disk) -- requiring 0.72\% of the model parameters.

\subsection{Dynamic Submodel Loading and On-the-fly Inference}
\label{sec:onthefly}
Having a well performing Submodel of a small size encourages us to propose the following modular approach to support large number of personalized models during inference. We first build our many-to-one (speaker-independent) Basemodel with a disabled Submodel using a residual factor of zero. This results into a fully optimized model that can be loaded directly into an accelerator to perform the inference fast and can be used to serve the general speaker-independent traffic. 

For a given request of speaker X, we load the corresponding speaker Submodel from fast storage, such as SSD. 
During inference, we feed the Submodel tensors to the model as extra inputs, overriding Basemodel adapter weights. 
Using the residual adapter-based Submodel, we have four tensors for each conformer block (layer norm stats, up and down projection and a residual factor) and we have 17 of those. During inference, we set the residual factor to 1 to enable the Submodel.

Testing the loading time for this Submodel (4.6MB) from a remote SSD takes 18$\pm$5.3ms per request, evaluated on 100 requests. This latency is acceptable for most production needs, even without caching Submodels. If we were not to use a Submodel approach, on the other hand, loading a fully fine-tuned model per speaker takes about 18.3$\pm$15 seconds. This latency is typically not acceptable per request for most real-time applications. Note that caching such large models for all speakers won't scale either, due to the large RAM requirement. Caching Submodels is far more feasible on the other hand.


\subsection{Scaling Training Using Embedding}
\label{sec:onehot}


We have seen that training a small Submodel exerts high quality model personalization that can be loaded dynamically per request. Personalizing a model for every speaker poses another complexity in scaling the training for many users.  We propose two approaches to alleviate this scaling issue: (1) training in parallel by 
parameterizing the Submodel by a speaker-id and learning one-hot embedding, (2) parameterizing the Submodel by a speaker-id and learning a real embedding space.


{\bf One-hot-embedding: } Since the Submodel described above has only 1.2M parameters per user, we propose to learn $N$ speaker Submodels in parallel by introducing a one-hot-embedding lookup table, as shown in Figure~\ref{fig:parrotronWithEmbedding}.  Each entry in this table points to a separate Submodel for each speaker, using their speaker-id ($0$ to $N-1$).
To separate all tensors of the residual adapter, the down-projection matrix ($d_i \times  d_b$) becomes a 3D tensor of ($N \times  d_i \times  d_b$), similarly, up-project matrix becomes a 3D tensor and the layer norm stats become 2D tensor, where each row belongs to a separate speaker. Thus, modeling N speakers with a conformer encoder of 17 blocks requires $N \times 17$ adapter layers in total.
During training, we select samples at random from all $N$ speakers.

Our effective Submodel size is now $N \times 17$ plus the size of the one-hot-embedding lookup table. This larger Submodel will slow down the on-the-fly-loading which may limit scaling inference. However, a key advantage of using a one-hot-embedding and a frozen Basemodel is that speaker parameters are independent across speakers. Therefore, as a post training phase, we split this large Submodel into speaker-dependent 1.2M parameter Submodels and discard the one-hot-embedding matrix. 

Table~\ref{tab:finetuning_vs_Submodel} shows that training a single job with a fixed training budget (30K steps) actually shows slightly {\em better} average WER (15.17\%) than training them separately (15.27\%). According to a paired t-test, this difference is not statistically significant, resulting into a `free' $128\times$ training throughput efficiency.
It is worthwhile mentioning that while it is might be possible to find a more optimized learning schedule with a more conservative number of steps per speaker for each individual job, scheduling many jobs and initializing and loading the Basemodel in each job have significant overhead on training.

Finally, we compare our one-hot-embedding approach to a baseline where we pool all the 128 speakers together into a single Submodel ignoring the identity of the speaker from all samples. This {\em Pooling Submodel} is a single residual adapter that is {\em not} parameterized by a speaker-id trained on all data. We increase the capacity of this Submodel by making the bottleneck dimension $d_b$ 128 instead of 64. Note that since this model is shared across speakers, we can't make this Submodel too large otherwise on-the-fly loading won't be sufficiently fast for real-time application. We observe the Pooling Submodel trained with the same data achieves the worst WER when compared to all other adaptation techniques. As shown in Table~\ref{tab:finetuning_vs_Submodel}, both training separately and using one-hot-embedding obtain significantly better WER than Pooling Submodel (22.64\%).

This method is related to~\cite{multitaskAdapters} where 
adapters are trained for $N$ tasks in parallel with multi-task objective using `hard parameter sharing', using the same hidden layers for all tasks. One-hot-embedding, on the other hand, gives each task (i.e., user) its own set of parameters.

{\bf Real embedding:} Although one-hot-embedding enables us to train many Submodels in the same training job, there is still a scaling limit to this approach. For example, training 1024 speakers in the same job would require ~1.3 billion {\em extra} parameters 
which would be challenging to fit in RAM. 

Instead of dedicating a separate Submodel for every speaker, we evaluate the use of a shared set of residual adapters per layer for all speakers. We do that by replacing the one-hot embedding by a real-valued embedding. The learned embedding matrix weights the potential of the $M$ adapters, per encoder layer. Our Submodel now becomes a real-embedding matrix to project a given speaker-id ($0$ to $N-1$) to a small $M$ dimensional real-valued vector and $M$ residual adapters, replicated {\em for each} encoder layer. Unlike one-hot-embedding, this approach enables us to choose $M$ that is significantly smaller than the number of speakers. In all our experiments, we choose $M$ to be 8. The size of the Submodel is now 8 $\times$ 1.2M parameters (using $d_b=64$), as opposed to 128 $\times$ 1.2M parameters as discussed for one-hot-embedding, during training. 


Similar to the one-hot embedding approach, we train our real-embedding-based Submodel on all the 128 speakers while freezing the Basemodel. Comparing the real-valued embedding to the one-hot embedding, we observe that the real-valued embedding is inferior to the one-hot embedding achieving a significantly worse absolute WER difference of 3.4\% in average ($p <0.001$ according to paired t-test). Although this approach shows degradation in terms of quality at modeling those 128 speakers, unlike one-hot-embedding, the real embedding approach enables us to scale to significantly larger number $N$ of speakers while keeping $M$ small. $M$ is chosen to control the latency of on-the-fly loading while still are able to train a large number of speakers in a single training job. 

It is worthwhile mentioning that AdapterFusion, described in \cite{AdapterFusionPfeiffer2020}, has shown how to learn knowledge sharing across tasks, by introducing a fusion network learned in a second training phase to fuse N task adapters. While it is a valuable approach for a small number of tasks, it results into a model that may not fit in RAM for a large number of tasks; and tasks can't be separated, as discussed in Section~\ref{sec:onehot}, preventing dynamic loading.  

\begin{figure}[ht]
  \centering
  \vspace{-2mm}

  \includegraphics[width=.8\linewidth]{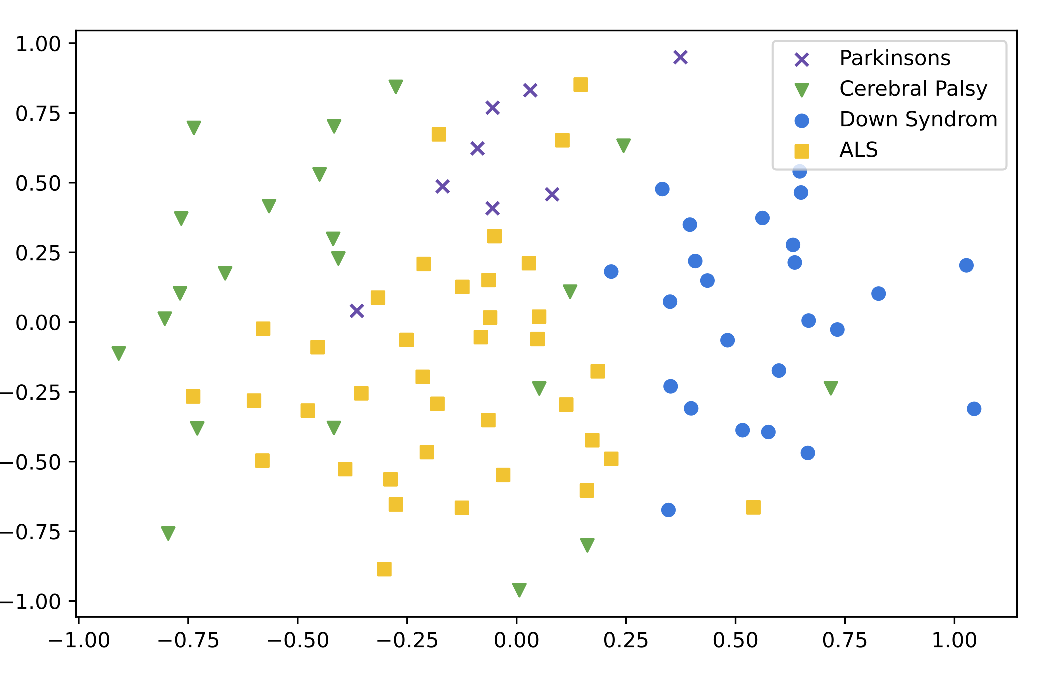}
  \caption{Embedding for Different Etiologies in 2D Space}
  \label{fig:embed2d}
  \vspace{-3mm}
\end{figure}

Another key advantage of real-embedding is sharing parameters across speakers.  To confirm that the model is, in fact, not memorizing those 128 speakers or learning a separate subspace for those speakers, we plot the embedding matrices for each speaker in 2D space. Concatenating all the 17 embedding 8D rows, we obtain 136D embedding vector per speaker, and then apply multi-dimensional-scaling~\cite{scikit-learn} to project the 136D embedding into 2D dimension (preserving distances), and plot the output of speakers from different etiologies in Figure~\ref{fig:embed2d}.

We observe that the learned embedding can easily separate speakers from different etiologies even after projection to 2D. Using the 136D vectors, and training a logistic classifier, we obtain an average pairwise accuracy of 92.8\%; ALS vs.~Down Syndrome is the most accurate classifier (98.5\% accuracy); and the least separable pair is ALS vs.~Parkinson's: 86.0\%. These results suggest that the Submodel is, in fact, not memorizing speakers but rather learns a structural correction of the Basemodel, otherwise points will  be randomly scattered in Figure~\ref{fig:embed2d}.

\vspace{-2mm}
\subsection{Embedding Submodel for new speakers with less data}

We hypothesize that learning an embedding space of atypical speech may be beneficial for new similar speakers with a {\em  limited} amount of personalization training data. To test this hypothesis, we make use of a new data set which consists of 10 new speakers, mutually exclusive from our 128 speakers. These 10 speakers have a variety of etiologies 
and levels of severity.

Using only 50 utterances for training and 200 held-out utterances for testing, we compare the following personalization approaches: (1) Adapt a residual-adapter-based Submodel from scratch for each user initialized at random (as shown in Section~\ref{sec:Submodel}); (2) Adapt the Pooling Submodel, which is trained on the pool of the 128 atypical speakers, described in Section~\ref{sec:onehot}; and (3) Adapt the real-embedding Submodel. We use the same number of model parameters in each approach.

For (3) since the embedding matrix is trained on the 128 speakers, to accommodate a new speaker, we discard the trained embedding matrix and replace it by a single 8D vector per layer, initialized randomly with a Gaussian noise of mean 0 and variance of 0.1 (indicating no prior knowledge about the speaker). During adaptation, however, we start from the pre-trained residual adapters of the real-embedding Submodel for each speaker. 
Adapting the pretrained real-embedding Submodel, while also learning the randomly initialized weighting embedding vector per speaker on {\em only} 50 utterances, achieves the best WER across 9 of 10 speakers (mean WER of 30.2\%).  Adapting an independent Submodel from scratch for each user obtains an average WER of 33.1\%. 
Adapting a Pooling Submodel, we observe the worst average WER of 44.3\%. 
Thus, starting from a model that shares parameters across speakers and enforces structure seems to be beneficial when personalization data is limited.


\section{Conclusions}
We propose a modular framework design to scale for specializing a large number of models for {\em real-time applications}, by creating a clear protocol
 and distinction between the roles
of {\em Basemodels} and {\em Submodels}.  We have defined a set of constraints a Submodel must obey to be used at scale during inference using dynamic on-the-fly loading strategy. We have also shown how independent Submodels can be trained in the same training job in parallel given the Basemodel. Demonstrating the effectiveness of this framework, we have tested our approaches for the task of speech model personalization of Parrotron, requiring a Submodel per speaker, to normalize and recognize atypical and dysarthric speech and showed $128\times$ training throughput with no loss in accuracy.  Learning a real-valued embedding space in our Submodels across speakers shows a great potential of scaling to an even larger number of speakers. We have discovered that the learned embedding linearly separates etiologies and can be used to reduce the amount of training data for new speakers.

\bibliographystyle{IEEEtran}

\bibliography{mybib}

\begin{thebibliography}{10}
\providecommand{\url}[1]{#1}
\csname url@samestyle\endcsname
\providecommand{\newblock}{\relax}
\providecommand{\bibinfo}[2]{#2}
\providecommand{\BIBentrySTDinterwordspacing}{\spaceskip=0pt\relax}
\providecommand{\BIBentryALTinterwordstretchfactor}{4}
\providecommand{\BIBentryALTinterwordspacing}{\spaceskip=\fontdimen2\font plus
\BIBentryALTinterwordstretchfactor\fontdimen3\font minus
  \fontdimen4\font\relax}
\providecommand{\BIBforeignlanguage}[2]{{%
\expandafter\ifx\csname l@#1\endcsname\relax
\typeout{** WARNING: IEEEtran.bst: No hyphenation pattern has been}%
\typeout{** loaded for the language `#1'. Using the pattern for}%
\typeout{** the default language instead.}%
\else
\language=\csname l@#1\endcsname
\fi
#2}}
\providecommand{\BIBdecl}{\relax}
\BIBdecl

\bibitem{Brown2020}
\BIBentryALTinterwordspacing
T.~B. Brown, B.~Mann, N.~Ryder, M.~Subbiah, J.~Kaplan, P.~Dhariwal,
  A.~Neelakantan, P.~Shyam, G.~Sastry, A.~Askell, S.~Agarwal,
  A.~Herbert{-}Voss, G.~Krueger, T.~Henighan, R.~Child, A.~Ramesh, D.~M.
  Ziegler, J.~Wu, C.~Winter, C.~Hesse, M.~Chen, E.~Sigler, M.~Litwin, S.~Gray,
  B.~Chess, J.~Clark, C.~Berner, S.~McCandlish, A.~Radford, I.~Sutskever, and
  D.~Amodei, ``Language models are few-shot learners,'' \emph{CoRR}, vol.
  abs/2005.14165, 2020. [Online]. Available:
  \url{https://arxiv.org/abs/2005.14165}
\BIBentrySTDinterwordspacing

\bibitem{Bommasani2021Fundation}
\BIBentryALTinterwordspacing
R.~B. et~al., ``On the opportunities and risks of foundation models,''
  \emph{CoRR}, vol. abs/2108.07258, 2021. [Online]. Available:
  \url{https://arxiv.org/abs/2108.07258}
\BIBentrySTDinterwordspacing

\bibitem{Shor2019}
J.~Shor, D.~Emanuel, O.~Lang, O.~Tuval, M.~Brenner, J.~Cattiau, F.~Vieira,
  M.~McNally, T.~Charbonneau, M.~Nollstadt, A.~Hassidim, and Y.~Matias,
  ``{Personalizing {ASR} for Dysarthric and Accented Speech with Limited
  Data},'' in \emph{Proc. Interspeech}, 2019, pp. 784--788.

\bibitem{biadsy2019parrotron}
F.~Biadsy, R.~J. Weiss, P.~J. Moreno, D.~Kanevsky, and Y.~Jia, ``Parrotron: An
  end-to-end speech-to-speech conversion model and its applications to
  hearing-impaired speech and speech separation,'' \emph{arXiv preprint
  arXiv:1904.04169}, 2019.

\bibitem{Gururangan2020}
\BIBentryALTinterwordspacing
S.~Gururangan, A.~Marasovic, S.~Swayamdipta, K.~Lo, I.~Beltagy, D.~Downey, and
  N.~A. Smith, ``Don't stop pretraining: Adapt language models to domains and
  tasks,'' \emph{CoRR}, vol. abs/2004.10964, 2020. [Online]. Available:
  \url{https://arxiv.org/abs/2004.10964}
\BIBentrySTDinterwordspacing

\bibitem{Yue2021}
\BIBentryALTinterwordspacing
C.~J. Reed, X.~Yue, A.~Nrusimha, S.~Ebrahimi, V.~Vijaykumar, R.~Mao, B.~Li,
  S.~Zhang, D.~Guillory, S.~Metzger, K.~Keutzer, and T.~Darrell,
  ``Self-supervised pretraining improves self-supervised pretraining,''
  \emph{CoRR}, vol. abs/2103.12718, 2021. [Online]. Available:
  \url{https://arxiv.org/abs/2103.12718}
\BIBentrySTDinterwordspacing

\bibitem{Shen2021LowRank}
\BIBentryALTinterwordspacing
E.~J. Hu, Y.~Shen, P.~Wallis, Z.~Allen{-}Zhu, Y.~Li, S.~Wang, and W.~Chen,
  ``Lora: Low-rank adaptation of large language models,'' \emph{CoRR}, vol.
  abs/2106.09685, 2021. [Online]. Available:
  \url{https://arxiv.org/abs/2106.09685}
\BIBentrySTDinterwordspacing

\bibitem{Zaken2019biases}
\BIBentryALTinterwordspacing
E.~B. Zaken, S.~Ravfogel, and Y.~Goldberg, ``Bitfit: Simple parameter-efficient
  fine-tuning for transformer-based masked language-models,'' \emph{CoRR}, vol.
  abs/2106.10199, 2021. [Online]. Available:
  \url{https://arxiv.org/abs/2106.10199}
\BIBentrySTDinterwordspacing

\bibitem{Lester2021}
\BIBentryALTinterwordspacing
B.~Lester, R.~Al{-}Rfou, and N.~Constant, ``The power of scale for
  parameter-efficient prompt tuning,'' \emph{CoRR}, vol. abs/2104.08691, 2021.
  [Online]. Available: \url{https://arxiv.org/abs/2104.08691}
\BIBentrySTDinterwordspacing

\bibitem{houlsby2019parameter}
N.~Houlsby, A.~Giurgiu, S.~Jastrzebski, B.~Morrone, Q.~De~Laroussilhe,
  A.~Gesmundo, M.~Attariyan, and S.~Gelly, ``Parameter-efficient transfer
  learning for {NLP},'' in \emph{Proc. ICML 2019}, 2019, pp. 2790--2799.

\bibitem{bapna2019simple}
A.~Bapna, N.~Arivazhagan, and O.~Firat, ``Simple, scalable adaptation for
  neural machine translation,'' \emph{arXiv preprint arXiv:1909.08478}, 2019.

\bibitem{tomanek2021residual}
K.~Tomanek, V.~Zayats, D.~Padfield, K.~Vaillancourt, and F.~Biadsy, ``Residual
  adapters for parameter-efficient {ASR} adaptation to atypical and accented
  speech,'' \emph{arXiv preprint arXiv:2109.06952}, 2021.

\bibitem{AdapterFusionPfeiffer2020}
\BIBentryALTinterwordspacing
J.~Pfeiffer, A.~Kamath, A.~R{\"{u}}ckl{\'{e}}, K.~Cho, and I.~Gurevych,
  ``Adapterfusion: Non-destructive task composition for transfer learning,''
  \emph{CoRR}, vol. abs/2005.00247, 2020. [Online]. Available:
  \url{https://arxiv.org/abs/2005.00247}
\BIBentrySTDinterwordspacing

\bibitem{multitaskAdapters}
\BIBentryALTinterwordspacing
A.~C. Stickland and I.~Murray, ``{BERT} and {PAL}s: Projected attention layers
  for efficient adaptation in multi-task learning,'' in \emph{Proceedings of
  the 36th International Conference on Machine Learning}, ser. Proceedings of
  Machine Learning Research, K.~Chaudhuri and R.~Salakhutdinov, Eds.,
  vol.~97.\hskip 1em plus 0.5em minus 0.4em\relax PMLR, 09--15 Jun 2019, pp.
  5986--5995. [Online]. Available:
  \url{https://proceedings.mlr.press/v97/stickland19a.html}
\BIBentrySTDinterwordspacing

\bibitem{Zhu2019}
H.~Zhu, L.~Wang, P.~Zhang, and Y.~Yan, ``{Multi-Accent Adaptation Based on Gate
  Mechanism},'' in \emph{Proc. Interspeech 2019}, 2019, pp. 744--748.

\bibitem{Gale2019}
R.~Gale, L.~Chen, J.~Dolata, J.~van Santen, and M.~Asgari, ``{Improving {ASR}
  Systems for Children with Autism and Language Impairment Using Domain-Focused
  DNN Transfer Techniques},'' in \emph{Proc. Interspeech 2019}, 2019, pp.
  11--15.

\bibitem{mustafa2014}
M.~B. Mustafa, S.~S. Salim, N.~Mohamed, B.~Al-Qatab, and C.~E. Siong,
  ``Severity-based adaptation with limited data for {ASR} to aid dysarthric
  speakers,'' \emph{PLOS ONE}, vol.~9, no.~1, pp. 1--11, 01 2014.

\bibitem{chen2021conformer}
Z.~Chen, B.~Ramabhadran, F.~Biadsy, X.~Zhang, Y.~Chen, L.~Jiang, A.~Chu,
  R.~Doshi, and P.~J.~M. Mengibar, ``Conformer parrotron: a faster and stronger
  end-to-end speech conversion and recognition model for atypical speech,'' in
  \emph{Proc. Interspeech}, 2021.

\bibitem{green2021}
J.~E. Green, R.~L. MacDonald, P.-P. Jiang, J.~Cattiau, R.~Heywood, R.~Cave,
  K.~Seaver, M.~A. Ladewig, J.~Tobin, M.~P. Brenner, P.~C. Nelson, and
  K.~Tomanek, ``Automatic speech recognition of disordered speech: Personalized
  models now outperforming human listeners on short phrases,'' in \emph{Proc.
  Interspeech 2021}, 2021.

\bibitem{kannan2019large}
A.~Kannan, A.~Datta, T.~N. Sainath, E.~Weinstein, B.~Ramabhadran, Y.~Wu,
  A.~Bapna, Z.~Chen, and S.~Lee, ``Large-scale multilingual speech recognition
  with a streaming end-to-end model,'' \emph{arXiv preprint arXiv:1909.05330},
  2019.

\bibitem{french999128}
\BIBentryALTinterwordspacing
R.~M. French, ``Catastrophic forgetting in connectionist networks,''
  \emph{Trends in Cognitive Sciences}, vol.~3, no.~4, pp. 128--135, 1999.
  [Online]. Available:
  \url{https://www.sciencedirect.com/science/article/pii/S1364661399012942}
\BIBentrySTDinterwordspacing

\bibitem{Swietojanski2024}
P.~Swietojanski and S.~Renals, ``Learning hidden unit contributions for
  unsupervised speaker adaptation of neural network acoustic models,'' in
  \emph{2014 IEEE Spoken Language Technology Workshop (SLT)}, 2014, pp.
  171--176.

\bibitem{doshi2021extending}
R.~Doshi, Y.~Chen, L.~Jiang, X.~Zhang, F.~Biadsy, B.~Ramabhadran, F.~Chu,
  A.~Rosenberg, and P.~J. Moreno, ``Extending parrotron: An end-to-end, speech
  conversion and speech recognition model for atypical speech,'' in
  \emph{ICASSP 2021-2021 IEEE International Conference on Acoustics, Speech and
  Signal Processing (ICASSP)}.\hskip 1em plus 0.5em minus 0.4em\relax IEEE,
  2021, pp. 6988--6992.

\bibitem{macdonald2021disordered}
B.~MacDonald, P.-P. Jiang, J.~Cattiau, R.~Heywood, R.~Cave, K.~Seaver,
  M.~Ladewig, J.~Tobin, M.~Brenner, P.~Q. Nelson \emph{et~al.}, ``Disordered
  speech data collection: Lessons learned at 1 million utterances from project
  euphonia,'' in \emph{Proc. Interspeech}, 2021.

\bibitem{conformer}
A.~Gulati, C.-C. Chiu, J.~Qin, J.~Yu, N.~Parmar, R.~Pang, S.~Wang, W.~Han,
  Y.~Wu, Y.~Zhang, and Z.~Zhang, Eds., \emph{Conformer: Convolution-augmented
  Transformer for Speech Recognition}, 2020.

\bibitem{scikit-learn}
F.~Pedregosa, G.~Varoquaux, A.~Gramfort, V.~Michel, B.~Thirion, O.~Grisel,
  M.~Blondel, P.~Prettenhofer, R.~Weiss, V.~Dubourg, J.~Vanderplas, A.~Passos,
  D.~Cournapeau, M.~Brucher, M.~Perrot, and E.~Duchesnay, ``Scikit-learn:
  Machine learning in {P}ython,'' \emph{Journal of Machine Learning Research},
  vol.~12, pp. 2825--2830, 2011.

\end{thebibliography}

\end{document}